# AlphaFold2 for protein structure prediction: Best practices and critical analyses

Ragousandirane Radjasandirane 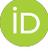 and Alexandre G. de Brevern*

Université Paris Cité and Université des Antilles and Université de la Réunion, BIGR, UMR_S1134, DSIMB Team, Inserm, F-75014 Paris, France.

Abstract

AlphaFold2 (AF2) has emerged in recent years as a groundbreaking innovation that has revolutionized several scientific fields, in particular structural biology, drug design and the elucidation of disease mechanisms. Many scientists now use AF2 on a daily basis, including non-specialist users. This chapter is aimed at the latter. Tips and tricks for getting the most out of AF2 to produce a high quality biological model are discussed here. We suggest to non-specialist users how to maintain a critical perspective when working with AF2 models and provide guidelines on how to properly evaluate them. After showing how to perform our own structure prediction using ColabFold, we list several ways to improve AF2 models by adding information that is missing from the original AF2 model. By using software such as AlphaFill to add cofactors and ligands to the models, or MODELLER to add disulfide bridges between cysteines, we guide users to build a high quality biological model suitable for applications such as drug design, protein interaction or molecular dynamics studies.

Running Title: AlphaFold2 for non-specialists.

1. Introduction

Understanding the three-dimensional structures of proteins is of major interest for comprehending their functions (*1*). Protein structures are involved in many essential biological processes, such as catalysis, binding, and signaling (*2*). Protein structures have been determined using experimental methods such as X-ray crystallography, but obtaining them is time-consuming and expensive. The paradigm that structure is better conserved than sequence has led to the development of a number of computational methods. Software-based approaches, which are far less expensive and time-consuming, have emerged as an alternative to experimental approaches for predicting the structure of proteins. Critical Assessment of Protein Structure Prediction (CASP) plays an important role in the evaluation and development of these approaches for protein structure prediction (*3*). Recently, some deep learning softwares have tremendously improved the prediction of protein structure (*4*) using new ways to effectively use the available data of protein structure in the Protein Data Bank (https://www.rcsb.org/). The most groundbreaking tool presented at CASP14 (*3*) is AlphaFold2 (AF2) (*5*), developed by DeepMind (a subsidiary of Alphabet Inc., Mountain View, CA, USA). AF2 is a deep learning tool that uses evolutionary information from Multiple Sequence Alignment (MSA) to capture dependencies and interactions between residues of a given protein sequence enabling accurate protein structure prediction; it has optimized a 21-million parameter model. During the CASP competition, prediction performances are evaluated using the Global Distance Test (GDT) metric (*6*) which measures the similarity between the predicted structure and the experimental one, ranging from 0 to 100. AF2 has achieved a GDT score greater than 90, meaning that it can produce structural models that are comparable to experimental ones (*3*). In comparison, the best performing method at CASP13 (which was the first version of AlphaFold) has reached a GDT score below 60 (*7*).

The success of AF2 has paved the way for the scientific community to do more research on protein structure prediction. One way of empowering the actual momentum of utilizing Artificial Intelligence (AI) for protein structure prediction is to facilitate the usage of these algorithms. First, the AlphaFold2 code was made available to the scientific community free of charge, but requires high local GPU power. Second, the ColabFold project (*8*) has provided an optimized and user-friendly free implementation of AlphaFold2, making it accessible for non specialist users.

AF2 is currently achieving excellent results in protein structure prediction, producing high quality models with often excellent accuracy. However, this field is constantly evolving, especially nowadays with the accelerated expansion of AI (*4*). Hence, several other alternatives have been released to predict the structure of proteins and try their best to match AF2 performances. One of the bottlenecks of AF2 computational time is the generation of MSA to capture evolutionary information of proteins. Therefore, Meta proposes their own algorithm named ESMFold (*9*) to predict the structure of proteins without using MSA which is a huge time-consuming step. ESMFold is far faster than AF2 with slightly poorer results.

Accurately predicting the structure of one protein is already a huge step forward to improve on many biological research fields like drug design, protein inhibition or protein engineering. Another significant improvement from AF2 is the prediction of protein complex models with AF2-Multimer released after the success of AF2 (*10*). AF2-Multimer is an extension of AF2 single chain prediction to multiple chain prediction by using an AF2-like pipeline in order to natively handle multimeric protein which produces state of the art accuracy for multimeric protein models.

AF2 has led to an increase in the number of high quality structural models produced, but as has been pointed out on several occasions, the question of proposing high quality models from the sequence is not yet feasible for all proteins especially transmembrane proteins (*11*). These difficulties arise from the inner flexibility or even disorder of certain regions, but also simply with the lack of experimental data for certain protein families. We must therefore always be precise and rigorous when analyzing the models generated.

In this chapter, we explore the best practices to use AlphaFold2 to predict protein structure. We also guide users in their critical analysis of AF2 models to ensure their biological relevance. Figure 1 summarizes the different points discussed in this chapter. From the sole knowledge of the protein sequence, access to structural information from the 3D structure or model can be done. Then, we guide the user to improve a model and produce a high quality biological model using software dedicated to add information on protein structure. The resulting model serves various applications, including drug design, the study of protein interactions within complexes, and Molecular Dynamics (MD) simulations to analyze the protein motion or behavior. This brief introduction on AF2 shows that major difficulty in using AF2 lies more in the query protein than with the method itself.

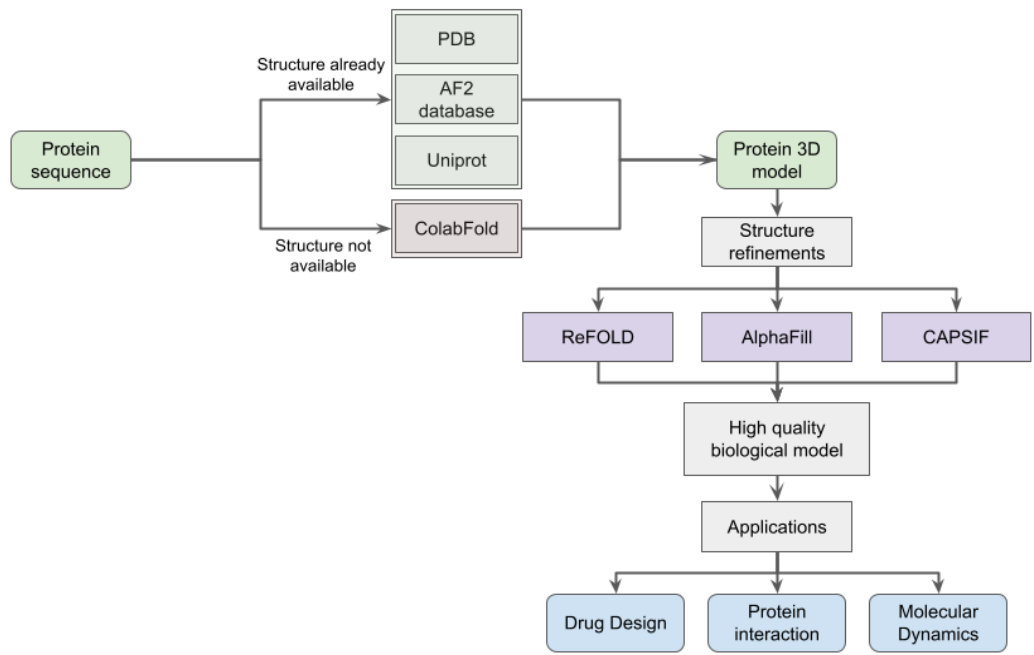

Figure 1. *General protocol to build and properly analyze a high quality biological model from sequence*.

2. Materials

This section presents the requirements for either running the original AF2 pipeline or to run ColabFold, the lighter, faster version of AF2 to predict protein structure. These requirements cover both technical specifications and the necessary understanding of the protein.

2.1. AlphaFold2 hardware requirements

All AlphaFold2 procedures, tools and code have been made available by DeepMind in a GitHub repository (https://github.com/google-deepmind/alphafold, accessed 27[th] September 2023). The whole package is free for both the academic and industrial communities. The installation documentation is of high quality and the number of commands required for installation is quite limited. Although running AlphaFold2 predictions does not require a cluster, you will still need a powerful and modern NVIDIA

GPU and 3TB of storage for the various databases to store locally. It is therefore not easy for all laboratories to install and use.

2.2. ColabFold, an easier alternative to AlphaFold2 pipeline

The AlphaFold2 code produced by DeepMind was perfectly suited to being made available in the form of GoogleColab. This system uses a Google form in which the user can perform all the actions necessary to run a programme and set its parameters. Several research groups have made their own Colab available, the best known being ColabFold (https://colab.research.google.com/github/sokrypton/ColabFold/blob/main/AlphaFold2.ipynb, accessed 27$^{th}$ September 2023) (8). All tools are available as web servers and do not require any installation by the user. It should be noted that there is a difference between a full installation and the use of AlphaFold2 and ColabFold. The latter uses more restricted databases than those available with the installation of AlphaFold2. This peculiarity is due to the fact that the computations are heavily offloaded on the host computer. However, it has very little effect on the quality of the results.

To use ColabFold, it is necessary to have a Google account in order to access Google server and computations ressources. Considering the large range of options made available this way, this requirement is a mild constraint.

2.3. Essential prerequisites about the query sequence

The complete amino acid sequence must be known. If the input is a nucleotide sequence, it must be edited to remove possible non coding regions and then translated. Importantly, only the mature protein can be processed. Although AlphaFold2 has led to a general improvement in results, another prerequisite is to know your protein well so as to avoid obvious mistakes, such as suggesting a transmembrane protein without a transmembrane domain.

3. Methods

Using AF2 to predict protein structure can be challenging as it requires special attention to certain points in order to build a correct biological model. We have detailed here the main steps of the proposed protocol by first obtaining a structural model which can be done using the AF2 database of predicted structure, especially if the structure of your protein has already been modeled, or execute AF2 through the ColabFold notebook for novel predictions. After interpreting the predicted structure using the pLDDT score to assess its prediction quality, the structure can be refined using optimization techniques to increase accuracy. Incorporation of cofactors or PTMs can further enhance the biological relevance of the model. We provide insights into the diverse applications of protein structures, highlighting their central role in areas such as drug design, functional analysis, and the exploration of protein-protein interactions.

### 3.1. Obtaining a structural model

Producing a structural model with AF2 can be highly expensive in computation time; this limitation can be explained by the AF2 pipeline. Indeed, AF2 uses Multiple Sequence Alignments (MSA), and MSAs generation by searching homologous sequences to the query could be a bottleneck of AF2. It uses HMMer and HHblits which are two sensitive homology detection methods to find homologous sequences to build the MSA. This search can take a very long time since current databases used by AF2 contain more than 2 TB of sequences.

Hence, multiple alternative solutions have been developed to allow a more user-friendly usage of AF2 and we will present some of them in the following. Two major scenarios exist: (i) you are (perhaps) lucky and your protein has already been modeled, or (ii) you must model yourself.

#### 3.1.1. Get protein structure from the Protein Data Bank

The first thing to do is to check whether the structure of your protein has already been solved and is available on the Protein Data Bank (*12*). If it is available, check that the sequence fully matches your sequence of interest (see Note 1).

#### 3.1.2. Search in already predicted structures on AlphaFold2 database

If not, DeepMind and EMBL-EBI have joined together to build a massive database of predicted protein structure using AF2. This database is freely accessible on https://alphafold.ebi.ac.uk/

(AlphaFold2 v4 release, accessed 27[th] September 2023) and contains more than 200 millions of models.

![Figure 2]

Figure 2. *Example queries on the EBI AlphaFold database*. (A) Search for terms related to calreticulin. (B) Result of this query. (C) Result of the query "calreticulin mouse".

You can search any keywords that would point toward a protein description, e.g. "E.coli", "receptors", "transmembrane protein", etc. You can choose to download only one protein of interest or download a subset of proteins. For example, you can download all protein models from *Arabidopsis thaliana* proteome (that would be 27.434 predicted structures) or all protein models from the Swiss-Prot (release 2023_05) database (542,378 predicted structures). Finally, if you are interested in downloading all of the 200 millions predicted structures, you need to be careful as it corresponds to 23 TB of disk storage. Figure 2 shows an example of a protein search in the EBI AlphaFold database. First, it is important to specify the search terms. A search for "calreticulin" gives more than 10 related terms, some very precise, others very broad (see Figure 2A). This query gives 3.414 results (see Figure 2B), far too many to analyze manually. Focusing on the mouse with the query "calreticulin mouse" returns only 14 results, including the first one searched for (see Figure 2C).

### 3.1.3. Retrieve protein structure predictions from UniProt

In a similar way, AF2 provides predicted structures for all proteins available in Swiss-Prot. You can retrieve them either from the AF2 database or directly in the Uniprot (https://www.uniprot.org, release 2023_05, accessed 27$^{th}$ September 2023) page of your protein. Figure 3 shows another way of looking for this muscle calreticulin protein. First of all, it is always a good idea to have a well-identified sequence. The NCBI site (*13*) (https://www.ncbi.nlm.nih.gov/, accessed 27$^{th}$ September 2023) is a classic portal to look for genes (see Figure 3A), and it will give you a direct result, even if your query is only approximate. It is even possible to download the sequence directly in FASTA format (see Figure 3B). However, it is more interesting to read the description of this gene (here named 12317, see Figure 3C). At the bottom of the form, we find the link to UniProt with the UniProt accession number P14211 (see Figure 3D). Conducting this query directly on UniProt yields 142 results, calreticulin comes second, the first result is calreticulin-3 (UniProt accession number Q9D9Q6). The UniProt files allow any disambiguation. The data sheet for mouse calreticulin tells us that domains of this protein already exist (see Figure 3E), but none of them is complete. The AF2 model is complete (see Figure 3F) but needs to be analyzed.

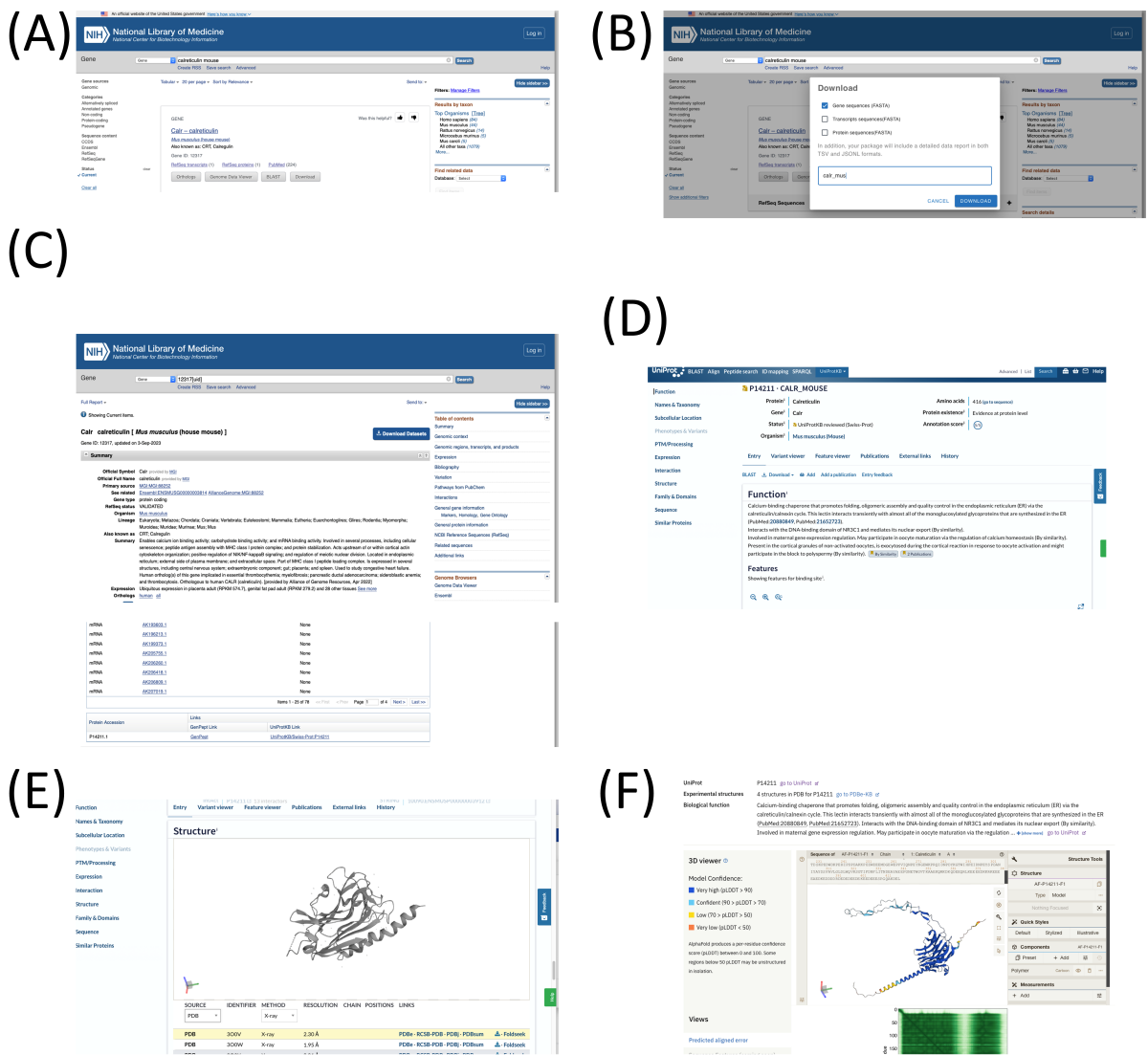

Figure 3. *Search AlphaFold model in UniProt*. (A) Search for the query "calreticulin mouse" on NCBI website. (B) corresponding sequence. (C) NCBI gene page with link to UniProt page. (D) UniProt page. (E) structure and (F) model of *mus musculus* calreticulin.

### 3.1.4. Protein structure prediction with ColabFold

If your protein is absent from the previously cited databases, you will have to run AF2 to predict the structure. In order to do this, ColabFold is a great alternative to the original AF2 pipeline (which requires a lot of computational resources). The ColabFold project has the aim of making structure prediction freely available to everyone. It is based on Google Colaboratory which is a Jupyter

Notebook hosted on Google servers. In this notebook, you have free access to important computational resources like GPUs (or TPUs) needed to run AF2. ColabFold is faster than AF2 since it uses MMseqs2 for fast sequence homology search instead of HHblits in the original AF2 pipeline. ColabFold is a user-friendly implementation of AF2 and requires only the sequence of the protein and nothing else from the user. You can access the notebook on https://colab.research.google.com/github/sokrypton/ColabFold/blob/main/AlphaFold2.ipynb (accessed 27$^{th}$ September 2023). In this notebook, you just have to put your sequence in the box "query_sequence", you can choose to use no template or the entire PDB (pdb100), which is the most efficient option. Then simply run all cells of the notebook by hitting "Run all" in "Runtime" (see Figure 4A and Note 2).

The number of parameters you can interact with is quite limited, which can be appreciable compared to other approaches like MODELLER (*14*) (https://salilab.org/modeller/, accessed 27$^{th}$ September 2023). You can choose to allow the use of template or doing the prediction without template information in the "template_mode" section (see Figure 4A). By default, ColabFold is running without using templates. The use of template improves the efficiency and accuracy of the prediction. It provides a starting point from an existing and well-defined protein structure, saving some computational time by not having to start the model from scratch. Moreover, the template provides crucial information about the possible structure of the protein to be predicted, which is particularly useful when the family of the protein is enriched in well-defined structures. However, there are some cases where using a template might not be beneficial. You may wish to predict the structure of your protein while minimizing potential bias (created by referring to existing protein structures as templates), as well as allowing the exploration of new and unexpected folds. It could also be interesting to not use templates if your protein is new or unusual.

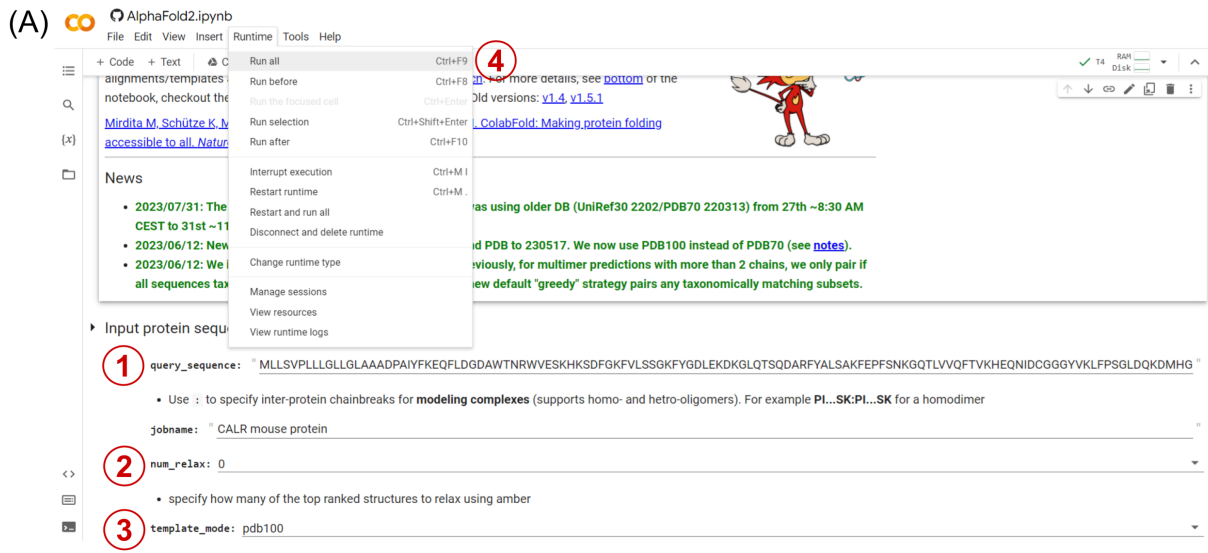

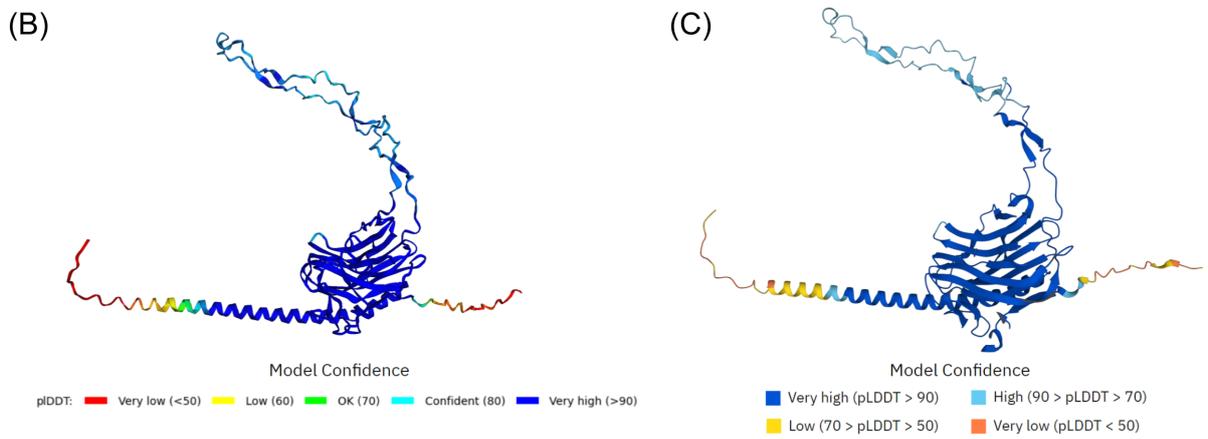

Figure 4. *Example run with ColabFold.* (A) Initialization of the inputs. Fill the sequence field in A1, choose if the prediction should use template or not in A2, then click on "Run all" in A3 to run the prediction. Prediction has five successive steps that can be run manually one by one. Please note that depending on the browser, the webpage could be slightly different. (B) Predicted structure for calreticulin mouse using ColabFold. (C) Predicted structure for calreticulin mouse from AF2 database.

In addition to the use of templates, it is possible to play with the number of top ranked predicted structures relaxations (0 by default, one or five can be set). This option uses an amber force field (*15*) to optimize protein contacts with very small molecular dynamics to "relax" the overall contacts of the structure. This can improve the global quality of models but will add more computational time to the prediction. More complex settings are available in the "Advanced settings" section, but are mainly for specialists (see Note 3).

The prediction process required 20 minutes for a 400 residue protein. When comparing predicted structure from ColabFold (Fig 4B) and from AF2 database (Fig 4C), slight differences can be seen especially at extremities of the protein (when there is no template available for this region of the protein) on which native AF2 performs better. Nevertheless, you can still have a very high quality protein structure in 20 minutes using ColabFold, particularly if there is a high quality template available for your protein.

### 3.2. Interpretation of the model

It is crucial to interpret the results of predictions and assess its quality and relevance to ensure its reliability and usability. To do so, AF2 provides some useful information as the pLDDT score which allows us to understand if the structure is well modeled or not, at the residue level. If the pLDDT confidence scores are all very good, interpretation of the model is quite straightforward. However, if they are not, further investigation is needed to determine whether the low pLDDT scores simply correspond to disordered regions or whether the problem is more complex.

#### 3.2.1. Evaluate quality prediction using the pLDDT score

AF2 gives a per residue score, namely the predicted local-distance difference test (pLDDT). This confidence score is an essential indicator of the local quality and global reliability of the predicted model. Generally, residues with a pLDDT exceeding 90 are considered as well predicted. Residues with low pLDDT (below 50) are considered as not reliable; they must be handled with care and need further analysis.

#### 3.2.2. Understanding regions with low pLDDT score

The low pLDDT scores can be explained by different reasons. First, AF2 is trained on experimental structures available in the PDB (*5*). Since the PDB is mainly composed of crystal structure (around

85% from PDB statistics), AF2 will struggle to model a region that is difficult to crystallize because this region will be under-represented in the PDB. This is the case for Intrinsically Disordered Regions (IDRs) or highly flexible regions. As these regions do not have a well-defined structure, they are therefore difficult to crystallize and not well represented in the PDB (*16*). Second, there are other cases of protein under-represented in the PDB, for example transmembrane proteins. These proteins are challenging to crystallize since the protein is embedded in a membrane which complexifies the process of protein extraction, purification and crystallization. Hence, AF2 will struggle to predict the structure of transmembrane proteins.

### 3.2.3. Predictions of regions with limited informations

Additionally to have a very low pLDDT, certain regions can be weirdly predicted by AF2. Indeed, regions with almost no information in AF2 knowledge can be modeled by long filaments surrounding the protein. These regions have undefined structure according to AF2 knowledge scope. It is the case for the intra cellular region of most transmembrane proteins for example. One way to know if the region is more likely to be disordered or if it is just a prediction with low information from AF2 is to use a disorder prediction methodology such as the Disopred tool (https://bio.tools/disopred3, accessed 28$^{th}$ September 2023) (*17*). Disopred is a machine-learning based approach to detect intrinsically disordered regions and can be used to interpret models from AF2. Therefore, if the region is not considered disordered, but only has low pLDDT values, more work will be required (see end of the chapter).

### 3.3. Improve the model

Although AF2 can produce models of very good quality (similar to experimental ones), predicted structures can still be improved since AF2 gives only a naked structure. Therefore, the model can be improved by adding some biological compound or by refining contacts within the structure.

### 3.3.1. Fine tune the parameters to get the best prediction possible

For advanced users, it is common to (i) play with the parameters of the advanced section of ColabFold and (ii) analyze in more detail the different alternative models proposed (maximum number of five). The principle is often the comparison of models in order to distinguish structurally similar from divergent parts of the protein. It is therefore sometimes interesting to rework these more complex parts, starting from their sequences alone. The process here is more complex and requires expertise and time.

### 3.3.2. Structural refinement using ReFOLD

There are several cases where AF2 models are not entirely reliable and can be improved. To address this issue, the use of 3D refinement tools can be useful. ReFOLD (*18*) is a refinement method that refines AF2 predictions without altering the model accuracy. The tool is available as a web server at https://www.reading.ac.uk/bioinf/ReFOLD/, accessed 28[th] September 2023). Authors of ReFOLD showed that 72% of models generated by ReFOLD taken from AF2 were improved.

### 3.3.3. Adding cofactors using AlphaFill

In a biological context, most proteins are bound to cofactors or ligands, but AF2 does not use them. AF2 gives only the naked protein structural model. The inclusion of cofactors and ligands helps the elucidation of both function and structural integrity. One way of adding these compounds to the predicted structure is to use additional tools specialized in this task in order to modify the model. AlphaFill (*19*) is one example of an algorithm adding ligand and cofactor to AF2 models. AlphaFill will search for structures that are homologous to the query one in the PDB by using BLAST (*20*). It will then filter the matches to keep only close homologs to the AF2 model, and add the compounds from these homologs to the AF2 model. You can choose to keep or not each compound based on the score given by AlphaFill. Similarly to the AF2 database, AlphaFill has its own database of refined AF2 models available at https://alphafill.eu/, accessed 28[th] September 2023). You can search a specific protein by its Uniprot accession number and find the refined model with added compounds. Alternatively, you can also upload your own AF2 model.

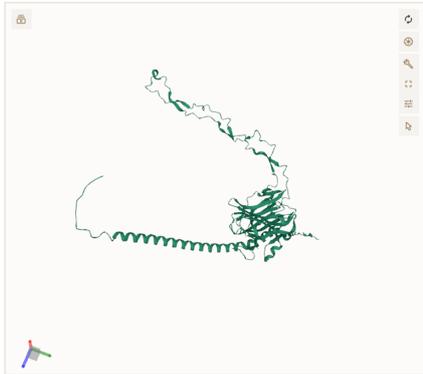
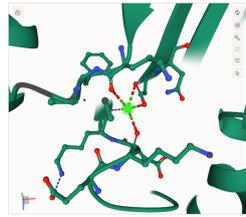
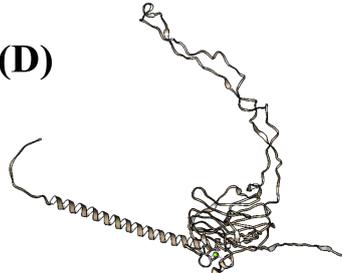

Figure 5. *Example of AlphaFill usage*. (A) First page, user put UniProt ID or a PDB file. (B) Using human calreticulin, P14211 is found. It is related to PDB id 3RGO chain A that has a $Ca^{2+}$ ion in its lectin domain. (C) Zoom on AlphaFill website and (D) refined PDB file visualized with Chimera software.

Figure 5 shows an example as before with the calreticulin protein. Using its UniProt accession number in AlphaFill (see Figure 5B), a calcium ion is detected in its central domain by analogy to

PDB ID 3RG0 chain A. It is highly logical as it is the resolved central domain of the protein (see Figure 5C). AlphaFill allows you to download the completed PDB files (see Figure 5D). Since AlphaFill works by analogy to existing PDB structures, it will not add information that is absent from the PDB. It is the case here as calreticulin binds half of $Ca^{2+}$ of endoplasmic reticulum and mainly on its terminal domain (*21*). These other $Ca^{2+}$ ions have not been added here by AlphaFill, as no PDB structures with these ions are currently available.

#### 3.3.4. Build disulfide bridges using MODELLER

AF2 does not explicitly predict the position of disulfide bridges, however, it can predict the correct angle of cysteine that are supposed to form a disulfide bridge. You can then add manually the disulfide bridge using software like MODELLER where you can specify where the disulfide bridge should be. To use MODELLER, you will need a template, which is the predicted structure from AF2, and the sequence of the protein. Then, you can run the MODELLER script with disulfide bridge constraint to produce a more relevant model (see Note 4). The use of MODELLER (*14*), although effective, is much less user-friendly than CollabFold. You have to (i) install it locally, (ii) obtain the license key (see Note 5), (iii) prepare the file explicitly requesting the disulfide bridge and (iv) generate a certain number of structural models with the creation of this (these) disulfide bridge(s).

#### 3.3.5. Introduce PTMs on the predicted structure

The AF2 predictions do not take into account consideration for post-translational modifications (PTMs), a critical aspect in understanding the functional complexity of proteins. To overcome this limitation, different resources can be employed to retrieve PTM positions. For example, phosphorylation information can be retrieved using Uniprot or Phosphosite plus (*23*) which are knowledge bases. Alternatively, PTM positions can be predicted using specialized software. These predictions can then be incorporated into molecular structures using dedicated tools. PTM prediction can be done using MusiteDeep software which uses sequence information to predict the positions of various PTMs, including phosphorylation, glycosylation, hydroxylation, and more (refer to (*22*) for a comprehensive list).

Some PTMs are relatively easy to model, such as phosphorylations, because they involve the addition of a single chemical group to a specific residue. Such PTM can be modeled using PyTMs (*23*), a PyMOL plugin that allows to easily to model ten common PTMs (e.g., phosphorylation, proline hydroxylation, methylation, acetylation, nitration, see (*23*) for a detailed list).

However, certain PTMs, such as glycosylation are more difficult to model as it requires (i) to locate the glycosylation site, (ii) to know which glycan is bound to the structure and (iii) then to introduce the glycan on the structure. All of these steps can be handled using the structure database GlycoShape (*24*) with its Re-Glyco tool. Given a UniProt accession, Re-Glyco extracts the AF2 predicted structure and glycosylation information from UniProt annotations. It then identifies the appropriate glycan for the protein structure and models it directly onto the AF2 structure (see Note 6). You can also upload your own PDB file. This tool significantly simplifies the modeling of glycosylation, allowing researchers to focus on more complex biological questions rather than technical skills needed to do such modeling.

### 3.3.6. Rebuild the model manually in cases of unreliable AF2 model

Finally, there will be the case where the AF2 model is just not reliable at all due to low prediction quality. It is important to note that only 1/3 of the human AF2 proteome is of atomistic quality and for 42% of these proteins, AF2 proposes models whose fold as such is uncertain (*25*).

It is then appropriate to try alternative ways of obtaining a structural model with (i) homologues search, (ii) sequence domain analysis such as PFAM (*26*), (iii) comparative modeling, (iv) *de novo* approaches, (v) other deep learning approaches, (vi) prediction of secondary structures, transmembrane segments, disorder, etc. It is therefore sometimes possible to find that a manually supervised model is more accurate than an AF2 model, Figure 6 (from AF2 Persepectives (*11*)) shows a human supervised vs. AF2 model of Scianna proteins. The AF2 model (Figure 6A) is incorrect because (i) it extends the membrane domain recklessly, (ii) it does not allow good positioning of the domains with a membrane, and (iii) it does not correspond to the biological knowledge of the different domains. The supervised model (Figure 6B), modeled domain by domain using the MODELLER software, is more in line with our current biological and topological knowledge.

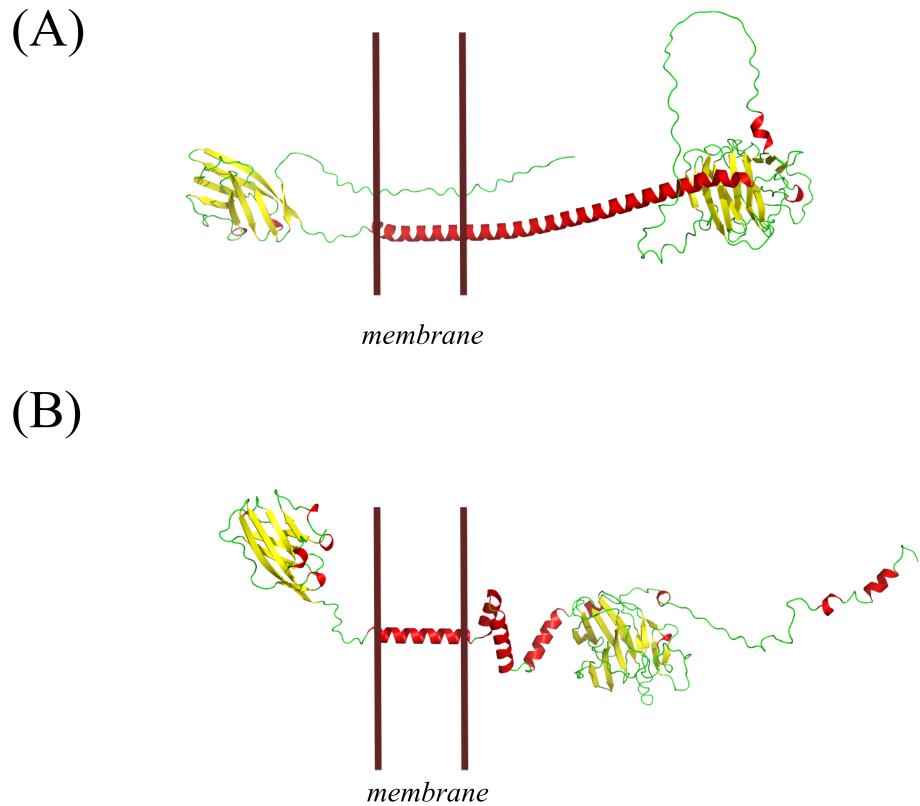

Figure 6. *Sometimes AF2 does not work*. (A) Model obtained from AF2; (B) Human-supervised model built with MODELLER from (*33*) adapted from (*11*). Membrane is positioned with two brown bars.

### 3.4. Application

Exploiting protein structures greatly enhances our understanding of fundamental processes in health and disease, and enables a multitude of applications which are detailed in this section. Protein structures play an important role in the developments of more effective vaccines and treatments, especially in the design of new drugs. The analysis of protein-protein interactions enables a better understanding of the complex relationships between proteins, providing valuable information for understanding disease mechanisms, identifying potential therapeutic targets, and optimizing drug development strategies. Protein dynamics can be analyzed using MD to assess the impact of mutations on the structure.

### 3.4.1. Designing new drugs with protein structure

Among the fields impacted by AF2 models is the design of new drugs (*27*). Drug design process is composed of several steps before reaching clinical tests (*28*). The first step is the identification and validation of the target protein with ideally a high quality 3D structure available on the PDB. If not, we can now use AF2 to produce a high quality structural model and rely on it to continue the process of drug discovery such as docking of drugs on the protein 3D model. Hence, AF2 has made drug discovery more achievable thanks to the availability of pertinent 3D models. A nice example is provided in (*27*).

### 3.4.2. Regions of interest of the structure

Having a 3D model of a protein is very useful to analyze its function or characteristics about the protein such as binding sites. Indeed, identification of potential regions of interest on the structure can be a strategic preparation step before diving into a project. For example, we can identify the active site of an enzyme and effectively design an experiment to analyze its behavior.

### 3.4.3. Protein-protein interaction and mutation effect analysis

Protein interactions can be analyzed thanks to AF2 models (*29*). AF2 can predict the structure of a multimer and therefore predict interfaces between chains of the multimer which can be useful to understand how chains (or proteins) interact and what are the residues involved in it. Protein-protein interactions are difficult to predict and AlphaFold-Multimer gives interesting results, but sometimes of low quality. Expertise is still required to analyze them (*30*). This information can be important and can lead to a better understanding of some disease where a patient has a mutation on residue involved in the interaction. In this context, DeepMind has released AlphaMissense (*31*) which is a modified version of AF2 that is specialized in predicting effects of mutations. It has made an additional forward step (but not a giant leap as AF2 did) over existing tools in the prediction of the effects of mutations. Further studies are needed to really try to understand these effects (*32*).

### 3.4.4. MD simulation analysis

One effective way to analyze a protein is by employing MD simulation using its 3D structure or model as a starting point. We can therefore apply MD to AF2 models to explore their dynamic behavior or study conformational changes which can give important insights to understand biological systems. One example of application can be the analysis of the impact of variants on the structure of the protein where the 3D models solely are not sufficient (*21*).

4. Notes

   1. *Protein sequence in the PDB.* If you find a PDB that matches your sequence, check carefully if the sequence in the PDB entry is exactly yours. Ofter, these sequences contain residue modifications that are engineered mutations. If so, reverse these mutations by doing point mutations using software like Pymol.

   2. *Google colab interface.* You may have a slightly different interface of Google colab where the Runtime field is not available or not working. In this case, you will need to run each cell manually by right clicking and hit "Ctrl + enter" for each of them.

   3. *Modifying parameters for advanced users.* When modifying parameters in ColabFold, take care on the purpose of each parameter in order to not produce a model that you don't understand or a wrong model.

   4. *MODELLER disulfide bridges.* If you want to model disulfide bridges with MODELLER, you will need a specific python code that you will find in the user manual (https://salilab.org/modeller/manual/node24.html, accessed 28$^{th}$ September 2023).

   5. *MODELLER license.* To get the license authorizing the use of MODELLER, you will need to fill a form in their website to be registered and to receive the license (https://salilab.org/modeller/registration.html, accessed 28$^{th}$ September 2023).

6. *Re-Glyco glycosylation information retrieving.* Re-Glyco facilitates glycosylation information retrieval by searching Uniprot annotations. However, users have the flexibility to manually input glycosylation details for each residue and select the appropriate glycan, using their knowledge of the protein. This functionality is available in the 'Advanced Glycosylation' section of Re-Glyco.